\documentclass[aps,prb,groupedaddress,showpacs,twocolumn]{revtex4}
\usepackage{graphicx}
\usepackage{amsmath}
\usepackage{bbm}
\usepackage{xspace}
\usepackage{color}

\newcommand{\se}{Sec.\@\xspace}

\newcommand{\ie}{i.\thinspace{}e.\@\xspace}

\newcommand{\ptl}{\partial}

\newcommand{\PDF}[2]{\frac{\ptl\, #1}{\ptl\, #2}}

\newcommand{\ve}[1]{{\bf #1}}
\newcommand{\mat}[1]{\mathsf{#1}}

\newcommand{\eq}[1]{Eq.\thinspace{}(\ref{#1})}

\newcommand{\eqqs}[2]{Eqs.\thinspace{}(\ref{#1}) and (\ref{#2})}

\newcommand{\fig}[1]{Fig.\thinspace{}\ref{#1}}

\newcommand{\fc}[1]{({#1})}
\newcommand{\figc}[2]{Fig.\thinspace{}\ref{#1}\thinspace{}\fc{#2}}

\newcommand{\Tr}{\mbox{Tr}}

\def\bra#1{\mathinner{\langle{#1}|}}
\def\ket#1{\mathinner{|{#1}\rangle}}

\begin{document}


\title{Spectral properties of strongly correlated bosons in two-dimensional optical lattices}


\author{Michael Knap}
\email[]{michael.knap@tugraz.at}
\affiliation{Institute of Theoretical and Computational Physics, Graz University of Technology, 8010 Graz, Austria}
\author{Enrico Arrigoni}
\affiliation{Institute of Theoretical and Computational Physics, Graz University of Technology, 8010 Graz, Austria}
\author{Wolfgang von der Linden}
\affiliation{Institute of Theoretical and Computational Physics, Graz University of Technology, 8010 Graz, Austria}


\date{\today}

\begin{abstract}
Spectral properties of the two-dimensional Bose-Hubbard model, which emulates ultracold gases of atoms confined in optical lattices, are investigated by means of the variational cluster approach. The phase boundary of the quantum phase transition from Mott phase to superfluid phase is calculated and compared to recent work. Moreover the single-particle spectral functions in both the first and the second Mott lobe are presented and the corresponding densities of states and momentum distributions are evaluated. A qualitatively similar intensity distribution of the spectral weight can be observed for spectral functions in the first and the second Mott lobe. 

\end{abstract}


\maketitle

\section{\label{sec:introduction}Introduction}
Pioneering experiments on ultracold gases of atoms trapped in optical lattices allowed for a direct observation of quantum many-body phenomena, such as the quantum phase transition from Mott phase to superfluid phase.\cite{jaksch_cold_1998, greiner_quantum_2002} Optical lattices are realized by counterpropagating laser beams, which form a periodic potential.\cite{bloch_many-body_2008} The bosonic particles located on the optical lattice gain kinetic energy when tunneling through the potential wells of neighboring sites of the periodic potential and they exhibit a repulsive interaction when a lattice site is occupied by more than one atom. A condensate of ultracold atoms can be driven from superfluid phase to Mott phase by gradually increasing the intensity of the laser beams. The potential wells of the optical lattice are shallow for low laser-beam intensity. Thus the bosonic particles can overcome the barrier easily and are delocalized on the whole lattice. However, for large intensity of the laser beams the potential wells are deep and there is little probability for the atoms to tunnel from one lattice site to another. This physical behavior can be described by the Bose-Hubbard (BH) model \cite{fisher_boson_1989} provided the gas of ultracold atoms is cooled such that only the lowest Bloch band of the periodic potential has to be taken into account.\cite{jaksch_cold_1998} The ground state of the BH model is superfluid when the local on-site repulsion between the atoms is small in comparison to the nearest-neighbor hopping strength whereas it is a Mott state for integer particle density and large on-site repulsion compared to the hopping strength. Due to these characteristics of the BH model the depth of the potential wells in optical lattices can be associated directly with the ratio of the on-site repulsion and the hopping strength. Ultracold atoms confined in optical lattices provide a very clean experimental realization of a strongly correlated many-body problem and the internal physical processes are well understood in comparison to conventional condensed-matter systems. There is large experimental control over the system parameters, such as the particle number, lattice size, and depth of the potential wells. Furthermore the sites of the optical lattice can be addressed individually due to the mesoscopic scale of the lattice.\cite{wurtz_experimental_2009}

The quantum phase transition from Mott phase to superfluid phase has
been first observed experimentally for ultracold rubidium atoms
trapped in a three-dimensional optical lattice
\cite{greiner_quantum_2002} and subsequently as well in optical
lattices of two dimensions.\cite{spielman_mott-insulator_2007,
  spielman_condensate_2008} 
The corresponding theoretical model, the two-dimensional (2D) BH model, has already been investigated to some detail in
literature. The phase diagram, which describes the quantum phase
transition from Mott phase to superfluid phase, has been investigated
thoroughly at the mean-field level (possibly including
Gaussian-fluctuation corrections).\cite{fisher_boson_1989,br.fa.93,ka.zi.93,
  sheshadri_superfluid_1993, van_oosten_quantum_2001,
  sachdev_quantum_2001, menotti_spectral_2008}
More accurate results  for the phase diagram 
from quantum Monte Carlo \cite{capogrosso-sansone_monte_2008} (QMC) simulations, 
variational approaches,\cite{ro.ko.91,ca.be.08}
 and strong-coupling perturbation 
theory\cite{fr.mo.94,freericks_strong-coupling_1996, elstner_dynamics_1999,bu.ve.05} 
are also available. 
The phase diagram for arbitrary integer fillings has been obtained
recently using the 
so-called diagrammatic process chain
approach.\cite{teichmann_bose-hubbard_2009,
  teichmann_process-chain_2009} Spectral functions of the two-dimensional BH model have been evaluated within a strong-coupling
approach \cite{sengupta_mott-insulator-to-superfluid_2005,
  elstner_dynamics_1999} and a variational mean field
approach.\cite{huber_dynamical_2007}

In the present paper we evaluate the border of the quantum phase transition from Mott phase to superfluid phase for the first two Mott lobes by means of the variational cluster approach (VCA),\cite{potthoff_variational_2003} 
and show that this method provides quite accurately the 
boundaries of the Mott phase, as compared with
more demanding QMC simulations and perturbative expansions.
In addition,  we study in detail the spectral
functions of the two-dimensional BH model in both the first and the second Mott
lobe,
which require computing the Green's function in real frequency domain. We also present the densities of states and momentum distributions corresponding to the spectral functions. %
Finally, as a technical point, we present an extension of 
the so-called 
$\mat{Q}$-matrix formalism, which has been originally proposed for
fermionic (anticommutator)
 Green's functions,\cite{aichhorn_variational_2006,zacher_evolution_2002} to
bosonic (commutator)
 Green's functions.\cite{aichhorn_quantum_2008} 
As we show below, this extension is nontrivial due to the nonunitary
nature of the Bogoliubov transformation for bosonic particles.

This paper is organized as follows. In \se~\ref{sec:model} the BH
model is introduced. Section \ref{sec:method} contains a short review
on the variational cluster approach and the extension of the $\mat{Q}$-matrix formalism. 
 Section \ref{sec:results} is devoted to
the spectral properties of the BH model in two dimensions. Here the phase diagram, spectral functions, densities of states, and momentum distributions are presented.
Finally, we summarize and conclude our
findings in \se~\ref{sec:conclusion}.

\section{\label{sec:model}Model}
The (grand-canonical)
Hamiltonian of the BH model \cite{fisher_boson_1989} is given by
\begin{equation}
 \hat{H}=-t \sum_{\left\langle i,\,j \right\rangle} \left( 
b_i^\dagger   \, b_j + \text{H.c.}\right)
+ \frac{U}{2} \sum_i \hat{n}_i\left(\hat{n}_i-1 \right) - \mu\,\hat{N}_p \; \mbox{,}
 \label{eq:bhm}
\end{equation} 
where $t$ is the nearest-neighbor hopping strength, $U$ is the local
on-site repulsion, and $\mu$ is the chemical potential. The angle brackets
in the first part of the Hamiltonian specify to sum over 
pairs of nearest
neighbors (each pair counted once).
 The operator $b_i^\dagger$ creates a particle at lattice site $i$ whereas $b_i$ annihilates a particle at site $i$. The total particle number 
\begin{equation}
 \hat{N}_p=\sum_i \hat{n}_i = \sum_i b_i^\dagger \, b_i
 \label{eq:bhm:n}
\end{equation} 
is conserved, since $[ \hat{H},\, \hat{N}_p ] = 0 $. The particles of the BH model are of bosonic character and thus the commutation relation $[ b_i,\,b_j^\dagger ] = \delta_{ij}$ is satisfied. The first term of the Hamiltonian models the hopping of a particle from lattice site $j$ to lattice site $i$. The second part describes the local on-site repulsion, which remains zero when a lattice site is unoccupied or occupied by only one particle. However, it increases proportional to $U$ for each additionally added particle. We consider the on-site repulsion $U$ as unit of energy. The third part of the Hamiltonian is necessary to perform calculations in grand-canonical ensemble, where the chemical potential $\mu$ controls the total particle number of the system. 

\section{\label{sec:method}Method}

\subsection{Variational cluster approach}

We use VCA\cite{potthoff_variational_2003} to evaluate the 
phase diagram and the
spectral functions of the 2D BH model. VCA is a variational extension
of the cluster perturbation theory \cite{snchal_spectral_2000,
  snchal_cluster_2002} and is based on the self-energy
functional approach (SFA) which has been originally proposed for
fermionic systems by
M. Potthoff.\cite{potthoff_self-energy-functional_2003-1,
  potthoff_self-energy-functional_2003} VCA has been extended to
bosonic systems as well.\cite{koller_variational_2006}

\begin{figure}
        \centering
        \includegraphics[width=0.48\textwidth]{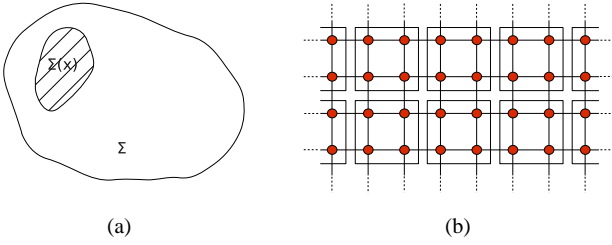}
        \caption{(Color online) \fc{a} The search space for the self-energy $\mat{\Sigma}$ is restricted to self-energies $\mat{\Sigma}(\mat{x})$ which are accessible via the reference system $\hat{H}^\prime$. \fc{b} Lattice decomposition of a square lattice into $2\times2$ site clusters. }
        \label{fig:num:vca}
\end{figure}

The SFA is based on the fact that Dyson's equation 
for the exact Green's function
is recovered at
the stationary point of the grand potential $\Omega[\mat{\Sigma}]$ 
considered as a functional of
the self-energy $\mat{\Sigma}$. Thus $\mat{\Sigma}$
corresponds, at the stationary point, to the real physical
self-energy. The self-energy functional $\Omega[\mat{\Sigma}]$ cannot be
evaluated directly as it contains the Legendre transform 
$F[\mat{\Sigma}]$ of the Luttinger-Ward functional.\cite{lu.wa.60,potthoff_self-energy-functional_2003-1}
However, 
the functional $F[\mat{\Sigma}]$ just depends on the interaction term of the
Hamiltonian, \ie, on the second term of \eq{eq:bhm}, and is thus
equivalent for all Hamiltonians which share a common interaction
part. Due to this property $F[\mat{\Sigma}]$ can be eliminated from the expression of the self-energy functional $\Omega[\mat{\Sigma}]$. For this purpose an exactly solvable, so-called ``reference,'' system $\hat{H}^\prime$ is constructed, which must be defined on the same lattice and must have the same interaction part as the original system $\hat{H}$. Thus both the self-energy functional of the original system $\Omega[\mat{\Sigma}]$ and the one of the reference system $\Omega^\prime[\mat{\Sigma}]$ contain the same $F[\mat{\Sigma}]$, which can be eliminated by comparison from the expressions of the two self-energy functionals. This yields for bosonic systems \cite{koller_variational_2006}
\begin{alignat}{2}
 \Omega[\mat{\Sigma}] &= \Omega^\prime[\mat{\Sigma}] &&- \Tr\,\ln(-(\mat{G}_0^{\prime\,-1}-\mat{\Sigma})) \nonumber \\
  & &&+ \Tr\,\ln(-(\mat{G}_0^{-1}-\mat{\Sigma})) \; \mbox{,}
 \label{eq:num:om5}
\end{alignat}
where quantities with prime correspond to the reference system and
$\mat{G}_0$ is the free Green's function. The free Green's function is
defined as $\mat{G}_0^{-1}\equiv(\omega+\mu)\hat{\mathbbm{1}} -
\mat{T}$, where $\mat{T}$ contains the hopping matrix and all other
one-particle parameters of the Hamiltonian except for the chemical
potential $\mu$, which is already treated separately in the
definition. The symbol $\Tr$ denotes a summation over bosonic
Matsubara frequencies and a trace over site indices. 
The self-energy functional 
$\Omega[\mat{\Sigma}]$ given by \eq{eq:num:om5} is exact. 
In
order to be able to evaluate  the functional, the search space of the
self-energy $\mat{\Sigma}$ has to be
restricted,\cite{potthoff_self-energy-functional_2003-1} 
which consists in an approximation.
More precisely,
the functional $\Omega[\mat{\Sigma}]$ is evaluated for
the 
subset of
self-energies 
available to
 the reference system $\hat{H}^\prime$, see
\figc{fig:num:vca}{a}. 
Practically, this is achieved by varying
the single-particle parameters of the reference Hamiltonian in order to find the
stationary point of the grand potential. Thus the functional
$\Omega[\mat{\Sigma}]$ becomes a function of the set $\mat{x}$ of
single-particle parameters of $\hat H'$
\begin{alignat}{2}
 \Omega(\mat{x}) &= \Omega^\prime(\mat{x}) &&- \Tr\,\ln(-(\mat{G}_0^{\prime\,-1}-\mat{\Sigma}(\mat{x}))) \nonumber \\
 & &&+ \Tr\,\ln(-(\mat{G}_0^{-1}-\mat{\Sigma}(\mat{x}))) \nonumber \\
 &= \Omega^\prime(\mat{x}) &&+ \Tr\,\ln(-\mat{G}^{\prime}(\mat{x})) -
 \Tr\,\ln(-\mat{G}(\mat{x}))  
\label{eqn:num:om5}
\end{alignat}
leading to the stationary condition 
\begin{equation}
 \PDF{\Omega (\mat{x})}{\mat{x}} = 0 \; \mbox{.}
 \label{eq:num:stat}
\end{equation}

In VCA the reference system is given by the decomposition of the total
system into identical clusters, see \figc{fig:num:vca}{b}. 
In order to implement the $\mat{Q}$-matrix approach, we solve each cluster
by means of the band Lanczos method.\cite{freund_roland_band_2000,
  aichhorn_variational_2006} 
The Green's function of the total system is obtained via the relation
\begin{equation}
 \mat{G}^{-1}(\omega) = \mat{G}^{\prime\,-1}(\omega) - \mat{V} \; \mbox{,}
 \label{eq:num:gVCA}
\end{equation}
which can be deduced from the Dyson equation of the total system
$\mat{G}^{-1} = \mat{G}_0^{-1} - \mat{\Sigma}(\mat{x})$ and the reference
system $\mat{G}^{\prime\,-1} = \mat{G}_0^{\prime\,-1} -
\mat{\Sigma}(\mat{x})$. 
The self-energy can be eliminated and it follows that
\[
 \mat{G}^{-1} = \mat{G}^{\prime\,-1} - (\mat{G}^{\prime\,-1}_0 - \mat{G}^{-1}_0)\;\mbox{.}
\]
The expression in parenthesis defines the matrix
\begin{align}
 \mat{V} &\equiv \mat{G}^{\prime\,-1}_0 - \mat{G}^{-1}_0 = ((\omega+\mu^\prime)\hat{\mathbbm{1}}-\mat{T}^\prime) - ((\omega+\mu)\hat{\mathbbm{1}}-\mat{T}) \nonumber\\
 &= -(\mu-\mu^\prime)\hat{\mathbbm{1}} + (\mat{T}-\mat{T}^\prime) \;\mbox{.}
\end{align}
With \eq{eq:num:gVCA} the grand potential $\Omega(\mat{x})$ can be rewritten as 
\begin{equation}
\label{om}
 \Omega(\mat{x}) = \Omega^\prime(\mat{x}) + \Tr \ln (\hat{\mathbbm{1}}-\mat{V}\mat{G}^\prime)\;\mbox{.}
\end{equation}

The decomposition of the $N$-site lattice into
clusters of  $L$ sites can be described
by  a superlattice. 
The original lattice is obtained by attaching a cluster to each site
of the superlattice.\cite{snchal_introduction_2008} 
A partial Fourier transform from superlattice indices to wave vectors
$\ve{\tilde{k}}$, which belong to the first Brillouin zone of the
superlattice, yields
 the total Green's function
\begin{equation}
 \mat{G}^{-1}(\ve{\tilde{k}},\, \omega) = \mat{G}^{\prime\,-1}(\omega) - \mat{V}(\ve{\tilde{k}}) \;\mbox{.}
 \label{eq:num:cpteqFourier}
\end{equation}
Due to the diagonality of $\mat{G}^\prime$ in the superlattice indices
its partial Fourier transform does not depend on $\ve{\tilde{k}}$. The
matrices in \eq{eq:num:cpteqFourier} are now defined in the space of
cluster-site indices and are thus of size $L\times L$. 
The $N$ wave vectors $\ve{k}$ from the Brillouin zone of the total lattice can be expressed as
\begin{equation}
 \ve{k}=\ve{\tilde{k}}+\ve{K}\;\mbox{,}
 \label{eqn:kDecomposition}
\end{equation}
where $\ve{K}$ belongs to both the reciprocal superlattice and the first Brillouin zone of the total lattice.\cite{snchal_introduction_2008}

\subsection{$\mat{Q}$-matrix formalism for bosonic systems}
The frequency integration implicit in the expression for
the grand potential, given in \eq{om}, can be carried out analytically,
yielding
at zero temperature\cite{koller_variational_2006,
  potthoff_self-energy-functional_2003, snchal_introduction_2008} 
\begin{equation}
 \Omega(\mat{x}) = \Omega^\prime(\mat{x}) + \sum_{\lambda_r^\prime<0} \lambda_r^\prime - \frac{1}{N_c} \sum_\ve{\tilde{k}} \sum_{\lambda_r ( \ve{\tilde{k}} )<0}  \lambda_r ( \ve{\tilde{k}}  ) \;\mbox{,} 
 \label{eqn:num:omIntZero}
\end{equation}
where $\lambda_r^\prime$ and $\lambda_r( \ve{\tilde{k}} )$ are the
poles of the cluster Green's function and total Green's
function, respectively. The number of clusters $N/L$ is denoted as
$N_c$. The poles $\lambda_r^\prime$ of the cluster Green's function  can
be readily obtained from the Lanczos method, whereas the poles of
the total Green's function $\lambda_r( \ve{\tilde{k}} )$ can be
evaluated with the 
so-called
$\mat{Q}$-matrix formalism, which was originally
 proposed for 
fermionic Green's
functions.\cite{aichhorn_variational_2006,zacher_evolution_2002} 
Here, we extend this formalism
to 
the generic case, \ie, we include bosonic
Green's functions. 
As we will see, this extension is nontrivial, since it involves
non-unitary transformations.

For zero temperature, the cluster Green's function
reads \cite{fetter.walecka} 
\begin{align}
 G_{ij}^\prime\left( \omega \right) &= \bra{\psi_0} a_i \frac{1}{\omega - ( \hat{H}^\prime - \omega_0 ) } a_j^\dagger \ket{\psi_0} \nonumber\\
 &- \epsilon \bra{\psi_0} a_j^\dagger \frac{1}{\omega + ( \hat{H}^\prime - \omega_0 ) } a_i \ket{\psi_0}\;\mbox{,}
 \label{num:gf}
\end{align}
where $\ket{\psi_0}$ is the ground state of the $N_p$ particle system,
$\omega_0$ is its (grand-canonical) energy, and $\epsilon=1$ ($\epsilon=-1$) for
bosonic (fermionic) Green's functions. The first 
term on the right-hand side of \eq{num:gf}
describes
single-particle excitations from the $N_p$ particle ground state
and can thus be referred to as particle term, whereas the second part corresponds to single-hole excitations and can be called hole term. 
Inserting the identity $\hat{\mathbbm{1}}=\sum_\gamma \ket{\gamma}\bra{\gamma}$ into
each part of \eq{num:gf}, where $\ket{\gamma}$ are the eigenvectors of
the reference Hamiltonian with corresponding eigenvalues $\omega^\prime_\gamma$,
yields the Lehmann representation of the Green's function
\begin{align}
 G'_{ij}\left( \omega \right) &= \sum_\alpha  \frac{\bra{\psi_0} a_i \ket{\alpha}\bra{\alpha} a_j^\dagger \ket{\psi_0} }{\omega - \left( \omega_\alpha^\prime - \omega_0 \right) } \nonumber \\
 &- \epsilon \sum_\beta \frac{\bra{\psi_0} a_j^\dagger \ket{\beta}\bra{\beta} a_i \ket{\psi_0} }{\omega + \left( \omega_\beta^\prime - \omega_0 \right) } \;\mbox{,}
 \label{num:gfLeh}
\end{align}
which can be cast into the form
\begin{equation}
 G'_{ij}\left( \omega \right) = \sum_\gamma Q_{i\gamma} \frac{1}{\omega - \lambda_\gamma^\prime} S_{\gamma \gamma} Q_{\gamma j}^\dagger \;\mbox{.}
 \label{num:gfQMatrix}
\end{equation}
In \eq{num:gfQMatrix}, we have introduced the following notation:
\begin{equation}
 Q_{\gamma j}^\dagger \equiv \left\lbrace  \begin{array}{ccc} \bra{\gamma} a_j^\dagger \ket{\psi_0} & \quad & \ket{\gamma} \in \mathcal{H}_{N_p+1} \\ \bra{\psi_0} a_j^\dagger \ket{\gamma} & \quad & \ket{\gamma} \in \mathcal{H}_{N_p-1} \end{array} \right. \;\mbox{,}
 \label{eq:num:qdef}
\end{equation}
\begin{equation}
 \lambda_{\gamma}^\prime \equiv \left\lbrace  \begin{array}{ccc} \omega_\gamma^\prime - \omega_0 & \quad & \ket{\gamma} \in \mathcal{H}_{N_p+1} \\ \omega_0-\omega_\gamma^\prime & \quad & \ket{\gamma} \in \mathcal{H}_{N_p-1} \end{array} \right. 
 \label{eq:num:lambdadef}
\end{equation}
and
\begin{equation}
 S_{\gamma \gamma^\prime} \equiv \left\lbrace  \begin{array}{ccl} \delta_{\gamma \gamma^\prime} & \quad & \ket{\gamma} \in \mathcal{H}_{N_p+1} \\ -\epsilon\,\delta_{\gamma \gamma^\prime} & \quad & \ket{\gamma} \in \mathcal{H}_{N_p-1}\end{array} \right.\;\mbox{,}
 \label{eq:num:Sdef}
\end{equation}
where $\mathcal{H}_{M}$ is the Hilbert space of an $M$ particle system.
With 
\begin{equation}
 g^\prime_{\gamma \gamma^\prime}\left( \omega \right)  \equiv \frac{\delta_{\gamma \gamma^\prime}}{\omega - \lambda_\gamma^\prime}
 \label{eq:num:gsmall}
\end{equation}
the cluster Green's function can be written in matrix notation
\begin{equation}
 \mat{G}^\prime \equiv \mat{Q} \, \mat{g}^\prime\left( \omega \right) \, \mat{S} \,\mat{Q}^\dagger \;\mbox{.}
 \label{num:gfShort}
\end{equation}
With the help of this expression
the VCA Green's function \eq{eq:num:cpteqFourier} can be rewritten as
\begin{align}
 \mat{G} & = \mat{G}^{\prime} \frac{1}{1-\mat{V}\,\mat{G}^{\prime}} = \mat{Q}\,\mat{g}^\prime\,\mat{S}\,\mat{Q}^\dagger \frac{1}{1-\mat{V}\,\mat{Q}\,\mat{g}^\prime\,\mat{S}\,\mat{Q}^\dagger} \nonumber \\
 &= \mat{Q}\,\mat{g}^\prime\,\mat{S}\,\mat{Q}^\dagger \left\lbrace 1 + \mat{V}\,\mat{Q}\,\mat{g}^\prime\,\mat{S}\,\mat{Q}^\dagger + \ldots \right\rbrace \nonumber \\
 &= \mat{Q}\,\mat{g}^\prime \left[ 1 - \mat{S}\,\mat{Q}^\dagger\,\mat{V}\,\mat{Q}\,\mat{g}^\prime \right]^{-1} \mat{S}\,\mat{Q}^\dagger \nonumber \\
 &= \mat{Q} \frac{1}{ \mat{g}^{\prime\,-1} - \mat{S}\,\mat{Q}^\dagger\,\mat{V}\,\mat{Q} } \mat{S}\,\mat{Q}^\dagger \;\mbox{,} 
 \label{eqn:num:gVCARewritten}
\end{align}
where
in the third step
 we expanded the fraction in a Taylor series. The matrix
 $\mat{g}^\prime$ is diagonal and contains the poles of the cluster
 Green's function $\mat{G}^\prime$, see \eq{eq:num:gsmall}. It can be
 written as $\mat{g}^{\prime\,-1}=\omega - \mat{\Lambda}$ with
 $\Lambda_{\gamma\gamma^\prime} = \lambda_\gamma^\prime \,
 \delta_{\gamma\gamma^\prime}$. Plugging this into 
\eq{eqn:num:gVCARewritten} 
yields
\begin{equation}
 \mat{G} = \mat{Q} \frac{1}{ \omega - (\mat{\Lambda} + \mat{S}\,\mat{Q}^\dagger\,\mat{V}\,\mat{Q}) } \mat{S}\,\mat{Q}^\dagger \;\mbox{.} 
 \label{eqn:num:GQmatrixLambda}
\end{equation}
We introduce the matrix $\mat{M} \equiv \mat{\Lambda} +
\mat{S}\,\mat{Q}^\dagger\,\mat{V}\,\mat{Q}$.
This matrix  can be diagonalized
as $\mat{M}\mat{X} = \mat{X}\mat{D}$, where $\mat{D}$ is a diagonal
matrix containing the eigenvalues of $\mat{M}$ and $\mat{X}$ is the
matrix of the eigenvectors of $\mat{M}$. The eigenvalue equation of
the matrix $\mat{M}$ can be rewritten as $\mat{M} =
\mat{X}\mat{D}\mat{X}^{-1}$, where $\mat{X}^{-1}$ is the inverse of $\mat{X}$ and not its transpose as $\mat{M}$ is a non-symmetric matrix. From that we obtain
\begin{equation}
 (\omega - \mat{M})^{-1} = \mat{X}(\omega - \mat{D})^{-1}\mat{X}^{-1}\;\mbox{.}
 \label{eqn:m}
\end{equation}
Therefore, the poles of the total Green's function $\mat{G}$ in
\eq{eqn:num:GQmatrixLambda} are the eigenvalues of the matrix
$\mat{M}$. The matrices $\mat {G}$ and $\mat{V}$ are defined on the
space of cluster-site indices. Thus $\mat {G}$ and $\mat{V}$ are of
size $L\times L$ and depend on the wave vector $\tilde{\ve{k}}$, see
\eq{eq:num:cpteqFourier}. The matrix $\mat{Q}$ is of size $L\times K$,
where $K$ is the dimension of the Krylov space generated in the band
Lanczos method. Due to the dependence of $\mat{V}$ on $\tilde{\ve{k}}$
the diagonalization of the matrix $\mat{M}$ yields $K$ eigenvalues
$D_{rr^\prime}=\lambda_r(\tilde{\ve{k}})\,\delta_{rr^\prime}$, which are used in
\eq{eqn:num:omIntZero}. The diagonalization has to be repeated for all
wave vectors $\tilde{\ve{k}}$. With that the grand potential
$\Omega(\mat{x})$ can be evaluated. 
The crucial point is that for bosonic Green's functions,
the entries of the
diagonal matrix $\mat{S}$ can be both $1$ as well as $-1$, see
\eq{eq:num:Sdef}. Therefore, the eigenvalue problem is not
symmetric.\cite{complexpoles} 

The factorization of the total lattice into clusters breaks the
translational symmetry of the lattice. Hence the total Green's function
would depend on two wave vectors $\ve{k}$ and $\ve{k}^\prime$, which
is certainly not correct for a periodic lattice. This has to be
circumvented by a periodization prescription that provides a total
Green's function $G(\ve{k},\,\omega)$ depending only on one wave vector
$\ve{k}$. The periodization prescription proposed in
Ref.~\onlinecite{snchal_spectral_2000}
 (Green's-function
periodization) reads as follows:
\begin{equation}
 G(\ve{k},\, \omega) = \frac{1}{L} \sum_{\alpha\beta} e^{-i\,\ve{k}\,\left( \ve{r}_\alpha - \ve{r}_\beta \right) } G_{\alpha\beta}( \ve{\tilde{k}},\, \omega )  \;\mbox{,}
 \label{eq:num:cpteqFourierFull}
\end{equation}
where $\ve{k}$ is a wave vector of the total lattice and
$\ve{r}_{\alpha}$ refers to lattice sites $\alpha$ of the
cluster. 
The wave vectors $\ve{\tilde{k}}$ in \eq{eq:num:cpteqFourierFull} can be replaced by the total wave vectors $\ve{k}$ as they just differ by a reciprocal superlattice wave vector, see \eq{eqn:kDecomposition}. With \eqqs{eqn:num:GQmatrixLambda}{eqn:m} the periodized Green's function can be rewritten in matrix notation
\begin{equation}
 G(\ve{k},\, \omega) = \ve{v}_\ve{k}^\dagger\,\mat{Q} \,\mat{X}\,(\omega - \mat{D})^{-1}\,\mat{X}^{-1} \, \mat{S}\,\mat{Q}^\dagger\,\ve{v}_\ve{k}\;\mbox{,}
 \label{eqn:greenfunctionMatrix}
\end{equation}
where the vector $\ve{v}_\ve{k}$ and its adjoint $\ve{v}_\ve{k}^\dagger$ contain $L$ plane waves 
\[\ve{v}_\ve{k}^\dagger \equiv \frac{1}{\sqrt{L}}\,\left( e^{-i\,\ve{k}\,\ve{r}_0},\,e^{-i\,\ve{k}\,\ve{r}_1},\,\ldots\,e^{-i\,\ve{k}\,\ve{r}_{L-1}}  \right) \;\mbox{.}\] 
There exists as well an alternative periodization
prescription where the self-energy $\mat{\Sigma}$ is
periodized.\cite{koller_variational_2006} This self-energy periodization should
prevent spurious gaps, which arise in the spectral function. 
However, at least for fermion systems, this procedure yields spurious metallic
bands in the Mott phase for arbitrarily large $U$.
Since we do not observe any spurious gaps in the spectral function of the
2D BH model we use the periodization on the Green's function defined in
\eq{eq:num:cpteqFourierFull}. 

With the wave-vector resolved Green's
function of the total system $G(\ve{k},\, \omega)$ we are able to
calculate the single-particle spectral function 
\begin{equation}
 A(\ve{k},\,\omega)\equiv-\frac{1}{\pi} \mbox{Im} \, G(\ve{k},\,\omega)\;\mbox{,}
 \label{eq:spe:spectralfunction}
\end{equation}
the density of states
\begin{equation}
 N(\omega)\equiv \int A(\ve{k},\,\omega) \, d\ve{k}  = \frac{1}{N} \sum_{\ve{k}} A(\ve{k},\,\omega)
 \label{eq:spe:dos}
\end{equation}
and the momentum distribution 
\[
 n(\ve{k})\equiv - \int_{-\infty}^0 A(\ve{k},\,\omega)\,d\omega \;\mbox{.}
\]
The frequency integration can be evaluated directly by means of the $\mat{Q}$-matrix formalism, which yields a sum of the residues of the Green's function, see \eq{eqn:greenfunctionMatrix}, corresponding to negative poles $\lambda_r(\ve{k})<0$,
\begin{equation}
 n(\ve{k})=\sum_{\lambda_r(\ve{k})<0} (\ve{v}_\ve{k}^\dagger\,\mat{Q} \,\mat{X})_r\,(\mat{X}^{-1} \, \mat{S}\,\mat{Q}^\dagger\,\ve{v}_\ve{k})_r \;\mbox{.}
 \label{eq:spe:nk}
\end{equation}

\section{\label{sec:results}Results}
The BH model exhibits a quantum phase transition from a Mott to a
superfluid phase when the ratio between the hopping strength and the
on-site repulsion $t/U$ is increased or when particles are added to or
removed from the system. The Mott phase is characterized by an integer particle density, a gap in the spectral function and zero compressibility.\cite{fisher_boson_1989} 

\begin{figure}
 \includegraphics[width=0.48\textwidth]{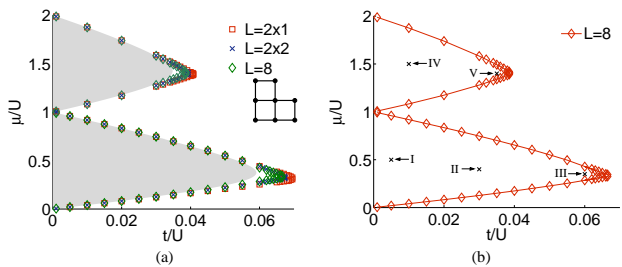}
 \caption{(Color online)
 Phase boundaries of the Mott phase of the 2D
   BH model (Mott lobes). \fc{a}
Results of our VCA calculation with 
 various cluster sizes for the reference system. The
   geometry of the 8-site cluster is visualized in the inset. The gray
   shaded area indicates the results of the process 
chain approach.\cite{teichmann_bose-hubbard_2009,teichmann_process-chain_2009}  
\fc{b} Phase boundaries obtained for
   the 8 site cluster. The marks refer to the parameters where we
   evaluated the spectral functions.\label{fig:pd}} 
\end{figure}
The first two Mott lobes of the 2D BH model obtained by means of VCA are shown in \fig{fig:pd}. We used the chemical potential $\mat{x}=\lbrace \mu \rbrace$ as variational parameter, which ensures a correct particle density of the total system.\cite{koller_variational_2006, aichhorn_antiferromagnetic_2006}
In contrast to the one-dimensional results \cite{koller_variational_2006,
  khner_one-dimensional_2000} the 
Mott lobes of the 2D BH model are round shaped. 
The gray shaded area in \figc{fig:pd}{a} presents the phase boundaries calculated
within the process chain approach 
 by {N.~Teichmann} \textit{et al.} in
 Refs.~\onlinecite{teichmann_bose-hubbard_2009} and \onlinecite{teichmann_process-chain_2009}, which are basically identical to the QMC results by B. {Capogrosso-Sansone} \textit{et al.}, see Ref.~\onlinecite{capogrosso-sansone_monte_2008}.
The agreement is quite good
for small hopping. However, VCA seems to  overestimate the critical value of the hopping $(t/U)_c$, which determines the tip of the Mott lobe. For the critical hopping of the first Mott lobe, we obtain approximately $(t/U)_c^1=0.067$ and for the second one $(t/U)_c^2=0.038$. Latest process chain approach ,\cite{teichmann_bose-hubbard_2009, teichmann_process-chain_2009} QMC (Ref.~\onlinecite{capogrosso-sansone_monte_2008}) and strong-coupling perturbation theory \cite{elstner_dynamics_1999} results yield $(t/U)_c^1=0.059$ and $(t/U)_c^2=0.035$ for the critical parameter of the first and second Mott lobe, respectively. 

\begin{figure}
\includegraphics[width=0.48\textwidth]{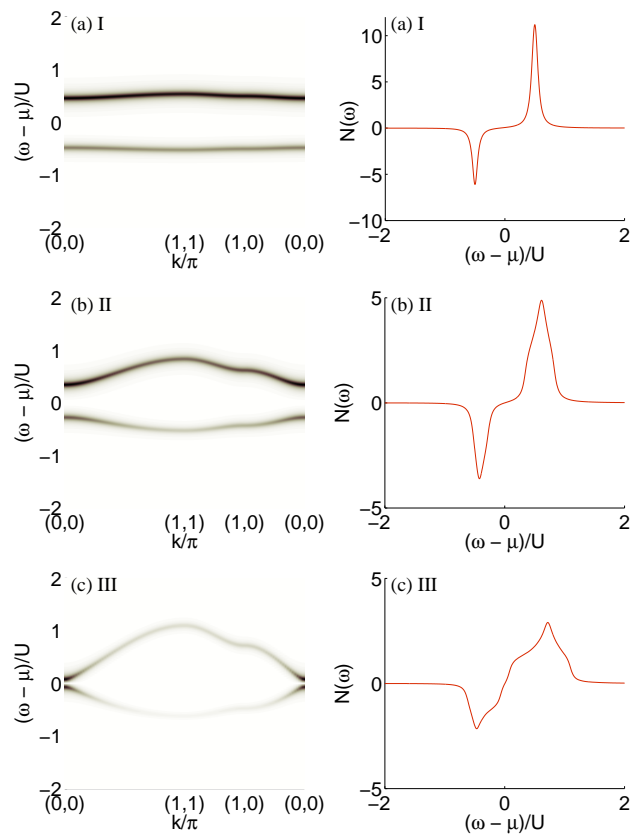}
\caption{(Color online) Spectral function $A(\ve{k},\,\omega)$, left column, and density of states $N(\omega)$, right column, in the first Mott lobe for the parameters \fc{a} $t/U=0.005$, $\mu/U=0.5$, \fc{b} $t/U=0.03$, $\mu/U=0.4$ and \fc{c} $t/U=0.06$, $\mu/U=0.35$. The captions of the subfigures refer to the marks in \figc{fig:pd}{b}.\label{fig:spectralLobe1}}
\end{figure}

The spectral functions $A(\ve{k},\,\omega)$ and the densities of states $N(\omega)$ for parameters of the first Mott lobe are shown in \fig{fig:spectralLobe1}.
The spectral function is displayed on the conventional path around the
Brillouin zone
$\ve{k}=(0,0)$ over $(\pi,\pi)$ to $(\pi,0)$ and back to $(0,0)$, and 
we use
an artificial imaginary-frequency broadening $\eta=0.05$. 
 A peculiarity of bosonic systems is that
the hole band of the spectral function has negative spectral weight
whereas the particle band has positive spectral weight. This follows
from the definition of the bosonic Green's function which has a
negative sign in front of the hole term, see \eq{num:gf}. In the
figures we always plot the absolute value of the spectral
function. 
The local density of states is defined as a wave-vector summation
of $A(\ve{k},\,\omega)$, see \eq{eq:spe:dos}. Therefore we observe a
negative peak in the density of states, which corresponds to the hole
band of the spectral function. 
For bosonic Green's functions the
density of states is not a probability distribution, as it contains
negative values. Taking the absolute value would yield an all positive
density of states, however, it would not be normed and is thus no
probability distribution either. 
For small hopping, the gap in the
spectral function is large and the bands are rather flat, \ie, the
width of the bands is small, see \fig{fig:spectralLobe1}. The
corresponding density of states contains two well-separated peaks. For
increasing hopping, the gap of the spectral function is decreasing and
the width of the bands is increasing. Pursuant to the spectral
function, the peaks in the density of states become broader for
increasing hopping. The intensity of the two bands is almost constant
for small hopping independent of the wave vector $\ve{k}$, whereas for
large hopping a large intensity can be observed at $\ve{k}=\ve{0}$. 

The boundaries of the Mott lobes correspond to the chemical
potential of the state with one additional particle (hole), which 
is obtained directly from
the 
single-particle (single-hole)
minimum excitation energy.
For this reason, we evaluate the phase diagram in Fig.~\ref{fig:pd}
by taking the minimal gap of the spectral function for each $t/U$,
which always occurs at $\ve{k}=\ve{0}$. 

\begin{figure}
\includegraphics[width=0.48\textwidth]{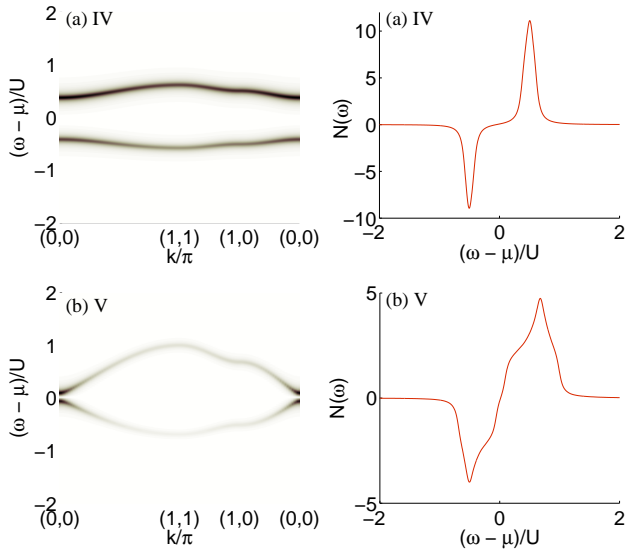}
 \caption{(Color online) Spectral function $A(\ve{k},\,\omega)$, left column, and density of states $N(\omega)$, right column, in the second Mott lobe for the parameters \fc{a} $t/U=0.01$, $\mu/U=1.5$ and \fc{b} $t/U=0.035$, $\mu/U=1.4$. The captions of the subfigures refer to the marks in \figc{fig:pd}{b}.\label{fig:spectralLobe2}}
\end{figure}

The spectral functions and densities of states in the second Mott lobe corresponding to the marks IV and V in \figc{fig:pd}{b} are shown in \fig{fig:spectralLobe2}.
Qualitatively they are very similar to the spectral functions and
densities of states in the first Mott lobe. Particularly, the
intensity distribution of the bands seem to be strongly related. Yet
the peaks of the density of states are larger 
due to the twice as large particle density within the second Mott lobe and thus the absolute value of the spectral weight in the second Mott lobe is larger than the one in the first Mott lobe. 

\begin{figure}
\includegraphics[width=0.48\textwidth]{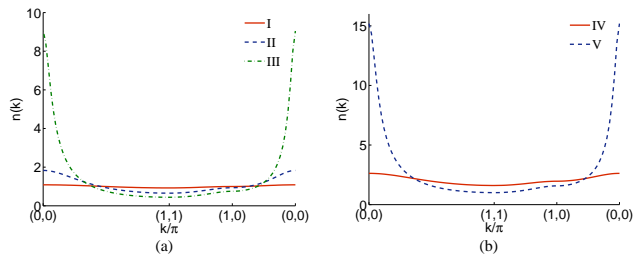}
 \caption{(Color online) 
Momentum distribution
 $n(\ve{k})$ in \fc{a} the first Mott lobe and \fc{b} the second Mott lobe. The Roman numerals in the legends refer to the parameters marked in \figc{fig:pd}{b}.\label{fig:nOfk}}
\end{figure}

\begin{figure}
\includegraphics[width=0.48\textwidth]{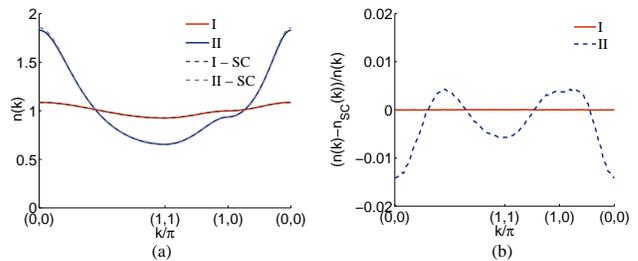}
 \caption{(Color online) 
Direct comparison (a) between the momentum distribution $n(\ve{k})$ obtained by means of VCA and a strong-coupling perturbation theory with scaling ansatz.\cite{freericks_strong-coupling_2009} The data identified with the letters SC correspond to the strong-coupling results. The relative deviations between VCA results and strong-coupling results with scaling ansatz are shown in (b). The Roman numerals in the legends refer to the parameters marked in \figc{fig:pd}{b}.\label{fig:nOfkComparison}}
\end{figure}

The momentum distribution
$n(\ve{k})$ corresponding to the spectral functions in the first and second Mott lobe are shown in \fig{fig:nOfk}. 
The particle density in the first Mott lobe is one and thus
$n(\ve{k})$ is centered around one in \figc{fig:nOfk}{a}. For the
second Mott lobe $n(\ve{k})$ is centered around two, see
\figc{fig:nOfk}{b}. 
The particle
density $n(\ve{k})$ is extremely flat for small hopping whereas it is
peaked at $\ve{k}=\ve{0}$ for large hopping, which is already a
precursor for the Bose-Einstein condensation where all particles
condense in the $\ve{k}=\ve{0}$ state. This behavior directly reflects the
intensity distribution of the bands in the spectral function.
There is excellent quantitative agreement between our VCA results for the momentum distribution and results obtained by means of QMC and a strong-coupling perturbation theory with scaling ansatz,\cite{freericks_strong-coupling_2009} see \fig{fig:nOfkComparison}.
We compare the momentum distributions for the parameters I
($t/U=0.005$) and II ($t/U=0.03$), and observe that the relative
deviations between our VCA results and 
an approach obtained by combining
strong-coupling
perturbation theory with a scaling ansatz~\cite{freericks_strong-coupling_2009}
are almost zero for
small hopping $t/U=0.005$ and less than one percent for medium hopping
$t/U=0.03$.
This latter methods is certainly more accurate than VCA in the evaluation
of the momentum distribution.
However, it should be mentioned that 
the information about the critical point (critical exponents and
critical hopping
  strength $(t/U)_c$) have to be inserted ``by hand,'' in order to
  optimize the results.
This information, in turn, must be extracted, e.~g., from a QMC calculation.
On the other hand, our VCA results
 are obtained directly without the need to introduce
  external parameters. 

\section{\label{sec:conclusion}Conclusions}
In the present paper, we presented and discussed results 
obtained
within the variational cluster approach 
for the spectral properties of the
two-dimensional
Bose-Hubbard
 Hamiltonian.
This is a minimal
model to describe 
bosonic ultracold atoms
confined in optical lattices,\cite{jaksch_cold_1998} 
and it
undergoes a
quantum phase transition from a Mott to a superfluid phase depending
on the chemical potential $\mu$, and the ratio between the hopping
strength and the on-site repulsion $t/U$. 
In particular,
 we determined the first two Mott lobes of the phase
diagram and found reasonable agreement with essentially exact results
from QMC simulations and from the process chain approach. 
In particular, 
the variational cluster approach yields very good results for the
phase boundaries apart from the region close to the lobe tip.
Here, 
strong-coupling expansions and QMC calculations 
are, clearly, much more accurate. 
Yet it should be emphasized that the computational effort is
considerably
 lower for VCA than for QMC.
Furthermore, we evaluated
spectral functions in the first and second Mott lobe. 
An important aspect of VCA is that the Green's function of the
  system is obtained directly in the real frequency domain, which
  allows for a direct calculation of the 
spectral function. On the
  other hand,  QMC quite generally 
provides correlation functions in imaginary time. Imaginary-time correlation functions have to be analytically continued to real frequencies,
which is a very ill-conditioned problem, as the data contain
statistical errors.
In QMC this analytical continuation is best carried out by means
of the maximum entropy method.
A very accurate dispersion (without spectral weight) has been also
obtained by a strong-coupling expansion.\cite{elstner_dynamics_1999}
The intensity
distribution of the spectral weight is similar for the spectral
functions of both Mott lobes, leading to an evenly distributed spectral
weight for small hopping strengths and to 
a distribution sharply peaked 
 at $ \ve{k}=\ve{0}$
for large hopping strengths. The latter
 indicates a precursor to  the Bose-Einstein condensation occurring
 above a certain critical hopping.
We also evaluated the densities of
states and momentum distributions corresponding to the calculated spectral
functions. 
We compared our VCA results for the momentum distribution with strong-coupling perturbation-theory results, where a scaling ansatz has been used, and found excellent quantitative agreement.
Finally, as a technical point, we extended the $\mat{Q}$-matrix formalism to
deal with bosonic Green's functions, which, in contrast to the fermionic case,
 produces a non-symmetric eigenvalue
problem.

\begin{acknowledgments}
We are
grateful to N.~Teichmann for providing the process chain approach data
of the phase diagram used in \fig{fig:pd}. We thank J.~K.~Freericks for sending us the self-consistently solved strong-coupling results with scaling ansatz shown in \fig{fig:nOfkComparison}. We made use of parts of the ALPS
library (Ref.~\onlinecite{albuquerque_alps_2007}) for the implementation of lattice geometries and for parameter parsing. 
M.K. wants to thank P.~Pippan for fruitful discussions. We acknowledge partial financial support from the Austrian Science Fund (FWF) under the doctoral program ``Numerical Simulations in
Technical Sciences'' No. W1208-N18 (M.K. and W.v.d.L.) and under project No. P18551-N16 (E.A.).
\end{acknowledgments}

%

\begin{thebibliography}{42}
\bibitem[{com()}]{complexpoles}
\bibinfo{note}{It can happen that some of the poles of the total Green's function
  \eq{eq:num:cpteqFourier} become complex. This is due to the fact that the
  matrix $\mat{M}$ in \eq{eqn:m} is not symmetric. For this reason, this
  anomaly can occur in the bosonic case only. With complex poles, the bosonic
  Green's function is no longer causal and, therefore, the variational solution
  is unphysical and must be discarded. This situation quite generally signals
  an instability towards another phase, such as  superfluidity.}
\end{thebibliography}

\end{document}